\documentclass[%
 reprint,
 amsmath,amssymb,
 aps,
 pra,
]{revtex4-1}

\usepackage{graphicx}
\usepackage{dcolumn}
\usepackage{bm}
\usepackage[dvipsnames]{xcolor}

\begin{document}


\title{Sub-ms, nondestructive, time-resolved quantum-state readout of a single, trapped neutral atom}

\author{Margaret E. Shea}
\email[corresponding author:~]{margaret.shea@alumni.duke.edu}
\affiliation{Department of Physics, Duke University, Durham, North Carolina 27708, USA}

\author{James A. Joseph}
\affiliation{Department of Electrical Engineering, Duke University, Durham, North Carolina 27708, USA}

\author{Paul M. Baker}
\affiliation{Department of Physics, Duke University, Durham, North Carolina 27708, USA}

\author{Jungsang Kim}
\affiliation{Department of Electrical Engineering, Duke University, Durham, North Carolina 27708, USA}

\author{Daniel J. Gauthier}
\affiliation{Department of Physics, The Ohio State University, Columbus, Ohio 43210 USA}


\date{\today}

\begin{abstract}
We achieve fast, nondestructive quantum-state readout via fluorescence detection of a single $^{87}$Rb atom in the 5$S_{1/2}$ ($F=2$) ground state held in an optical dipole trap. The atom is driven by linearly-polarized readout laser beams, making the scheme insensitive to the distribution of atomic population in the magnetic sub-levels. We demonstrate a readout fidelity of $97.6\pm0.2\%$ in a readout time of $160\pm20$ $\mu$s with the atom retained in $>97\%$ of the trials, representing an advancement over other magnetic-state-insensitive techniques. We demonstrate that the $F=2$ state is partially protected from optical pumping by the distribution of the dipole matrix elements for the various transitions and the AC-Stark shifts from the optical trap.  Our results are likely to find application in neutral-atom quantum computing and simulation.
\end{abstract}

\maketitle

Optically-trapped neutral-atom qubits have emerged as a promising platform for quantum computing, quantum simulation \cite{SaffmanProgressChallenges, 51atom, BrowIsing, LukinGates, EndresNature, SaffmanArray, LukinRydberg}, and the study of fundamental light-atom interactions \cite{BrowWig,KurtseiferHeat}. Most experiments require reading out the atom's internal quantum state to determine the outcome of a given protocol, such as in quantum computation.  It is desirable to perform the measurement with high fidelity in a short readout time without losing the atom from the trap so that the experiment can be repeated rapidly. This quantum-state readout requires balancing conflicting physical effects, such as heating from the fluorescent photons during readout and optical pumping of the atom to another state.    

Here we use linearly-polarized light to discriminate between the two hyperfine ground states of a single \textsuperscript{87}Rb atom via state-dependent fluorescence while the atom is held in an optical dipole trap (ODT).  We use a single high-numerical-aperture lens to both create the optical trap and collect the atomic fluorescence. Under optimized conditions, we achieve a discrimination fidelity of $>97$\% in a measurement time of $160$ $\mu s$, which can be used as a state-readout of a qubit register. In addition, by time-tagging the incoming photons and using a model of the readout protocol \cite{NoekOL,Crain}, we determine the decay rate of the fluorescent light and identify atom heating as a primary factor limiting the system performance.   Furthermore, we develop a rate-equation model for the fluorescence process \cite{Shea}, which reveals that the AC-Stark shifts help maximize the measurement fidelity. 

\begin{figure}
	\begin{center} \includegraphics[width=3.25 in]{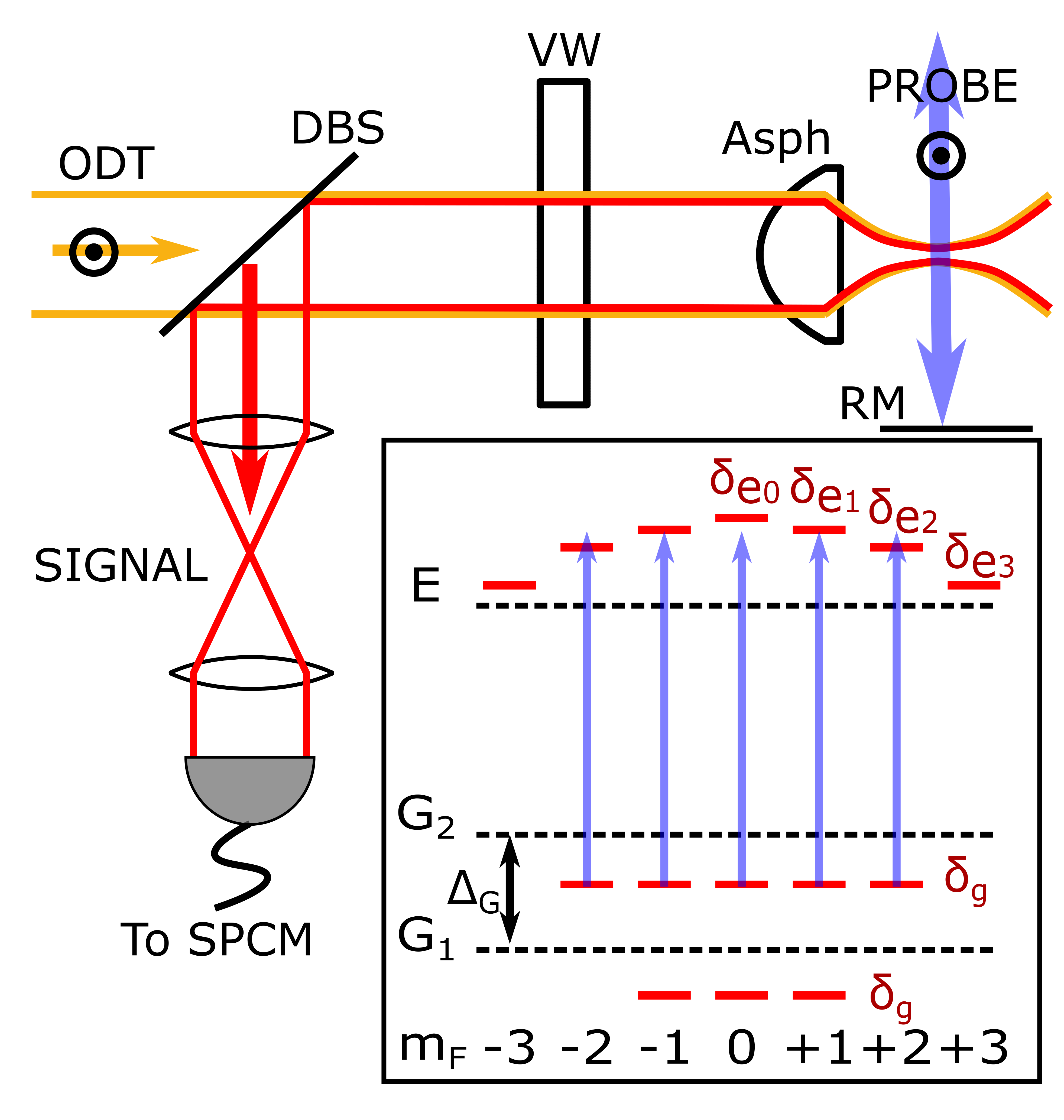} \end{center}
	\caption{The ODT light passes through the vacuum window (VW) and is focused by an asphere (Asph) to form the trap. The probe light is retroreflected by a mirror (RM) to produce counter-propagating beams. The signal is separated from the ODT light by a dichroic beamsplitter (DBS) before being directed to the SPCM. Inset: The atom is probed by the $\pi$-polarized probe light on the $G_{2}\rightarrow E$ transition. The energy levels, hyperfine splittings ($\Delta$) and AC-Stark shifts ($\delta$) are defined in the text.}\label{fig:setup}
\end{figure}

In our experiments, quantum-state readout requires measuring whether the $^{87}$Rb atom is in the $F=1$ or $F=2$ hyperfine levels of the $5$ $S_{1/2}$ ground state. These levels are shown in the inset of Fig. \ref{fig:setup} and are labeled $G_{1}$ and $G_{2}$, respectively (splitting $\Delta_G=2\pi$(6.8 GHz)). Quantum-state readout is achieved via fluorescence detection using $\pi$-polarized light that is nearly resonant with the $5S_{1/2}(F=2)\rightarrow 5P_{3/2}(F'=3)$ transition ($G_{2} \rightarrow E$)  \cite{Wineland80}. An atom in the $F=2$ ground state scatters photons when illuminated by the readout beam and appears bright during the measurement time (the `bright state'), while atoms in the $F=1$ ground state essentially do not fluoresce (the `dark state') due to the large detuning $\Delta_G$. The measurement decision as to whether the atom is bright or dark is based on recording a threshold number of photons $n_{thresh}$ \cite{Lucas}. This readout scheme does not require optically pumping the atom into a particular magnetic sub-level before the measurement. 

Previous studies have used fluorescence detection methods for nondestructive quantum-state readout of single atoms and arrays of trapped neutral atoms using circularly-polarized readout light \cite{BrowaeysND, SaffmanND, MeschedeND}, which requires pumping the atom into the $m_{F}=\pm F$  magnetic sub-level.  The optical pumping process is hampered by the AC-Stark shifts caused by the ODT and requires a pumping time on the order of a few ms or turning off the ODT.  Another study used linearly-polarized readout light \cite{ChapmanND}, but did not focus on minimizing the measurement time. We note that readout in 100 $\mu s$ has been reported in an atom-chip experiment that extracted an atom from a Bose Einstein Condensate and trapped it in a cavity \cite{Reichel}. This requires a substantially different experimental setup than the common free-space ODT used here. There has also been recent progress in quantum measurement using a Stern-Gerlach-type approach to infer the state by splitting atomic wavefunctions \cite{WeissSG}.

Our experimental setup is shown in Fig. \ref{fig:setup}.  A single $^{87}$Rb atom is confined in an 1.28-mK-deep ODT created by a linearly-polarized, 852-nm-wavelength, 40-mW-maximum-power beam focused to an estimated 3-$\mu$m-waist (1/$e$ intensity diameter) by an off-the-shelf, NA=0.54 asphere mounted in the vacuum chamber. The atom is loaded into the ODT from a magneto-optical trap (MOT).  Once loaded, the presence of the atom is verified using an atom detection sequence that cools the atom during detection similar to that reported in \cite{ShihPRA}. The quantum state of the atom is readout using counter-propagating laser beams linearly polarized along the same axis as the ODT beam. No external magnetic field is applied, greatly simplifying the experimental setup and required alignment. The quantization axis is defined along the direction of the linear-polarization of the trapping beam. In the collection path, the atomic fluorescence is focused to an intermediate plane where a spatial filter reduces background scatter. The atomic fluorescence is coupled into a multimode fiber that directs it to a single photon counting module (SPCM) avalanche photodiode from Perkin-Elmer (part number AQR14+) with a dark count rate of 150 Hz. The multimode fiber acts as a secondary spatial filter to further reduce background counts.  

The $\pi$-polarized ODT light shifts the magnetic sublevels of the atom due to the AC-Stark effect \cite{ShihPRA, PRAComment}, as shown in  the inset of Fig. \ref{fig:setup}. The ground state magnetic sublevels uniformly shift by $\delta_{g}=-27$ MHz, creating the trapping potential. The effect lifts the degeneracy of the $F'=3$ ($E$) excited state with shifts of $\delta_{e0}=21$ MHz, $\delta_{e1}=19$ MHz, $\delta_{e2}=13$ MHz, and $\delta_{e3}=3$ MHz, yielding different resonance frequencies for each $\Delta m_{F}=0$ transition probed by the linearly-polarized readout light. This causes a broadening of the atom's natural linewidth for $\pi$-polarized readout light tuned to the $F=2\rightarrow F'=3$ transition \cite{ShihPRA}. We observe a broadened linewidth of $\sim 13$ MHz and find that the atomic fluorescence is maximized when the readout-beam frequency is detuned $+ 46$ MHz from the untrapped atom's resonance (to the high-frequency side of the resonance), a frequency weighted towards the $m_{F}=0$ transition frequency. This observation is consistent with the fact that the $m_{F}=0$ transition has the largest Clebsch-Gordan coefficient and the population tends to accrue in the $m_{F}=0$ state, as discussed below.

To determine the fidelity of the quantum-state readout protocol, the atom is first prepared into either the bright state ($F=2$) by pumping the atom for 100 $\mu$s using the MOT repump beams, or the dark state ($F=1$) by pumping the atom for 5 ms using the MOT cooling beams. The atom's state is measured using a single 200-$\mu$s-long readout pulse. If more than $n_{thresh}$ photons are detected during the readout time, the atom is declared to be in the bright state ($F=2$). If fewer than $n_{thresh}$ photons are detected, the atom is declared to be in the dark state ($F=1$). Once the readout is complete, another atom detection sequence is performed to verify that the atom remains in the trap and that the readout is nondestructive. 

The readout fidelity is determined using the relation  $\mathcal{F}_{n_{thresh}}=1-(\epsilon_{B}+\epsilon_{D})/2$, where $\epsilon_{B}$ is the bright-state error, $\epsilon_{D}$ is the dark-state error, and $n_{thresh}$ is the number of photons needed to classify the atom as bright or dark. The bright-state error is the probability that an atom prepared in the bright-state is detected as dark. In other words, for a set of experiments where the atom is prepared in the bright state, the bright-state error is the fraction of measurements in which fewer than $n_{thresh}$ photons is detected. Likewise, the dark-state error is the probability that an atom prepared in the dark state is detected as bright. These errors include both detection and preparation errors. By time-tagging the photon detection time relative to the start of the readout pulse, we can reconstruct the state readout fidelity during each 1 $\mu$s interval of the 200-$\mu$s-long readout time and for any photon number threshold, a measurement not previously reported for neutral atoms. 

\begin{figure}
	\begin{center} \includegraphics[width=3.25 in]{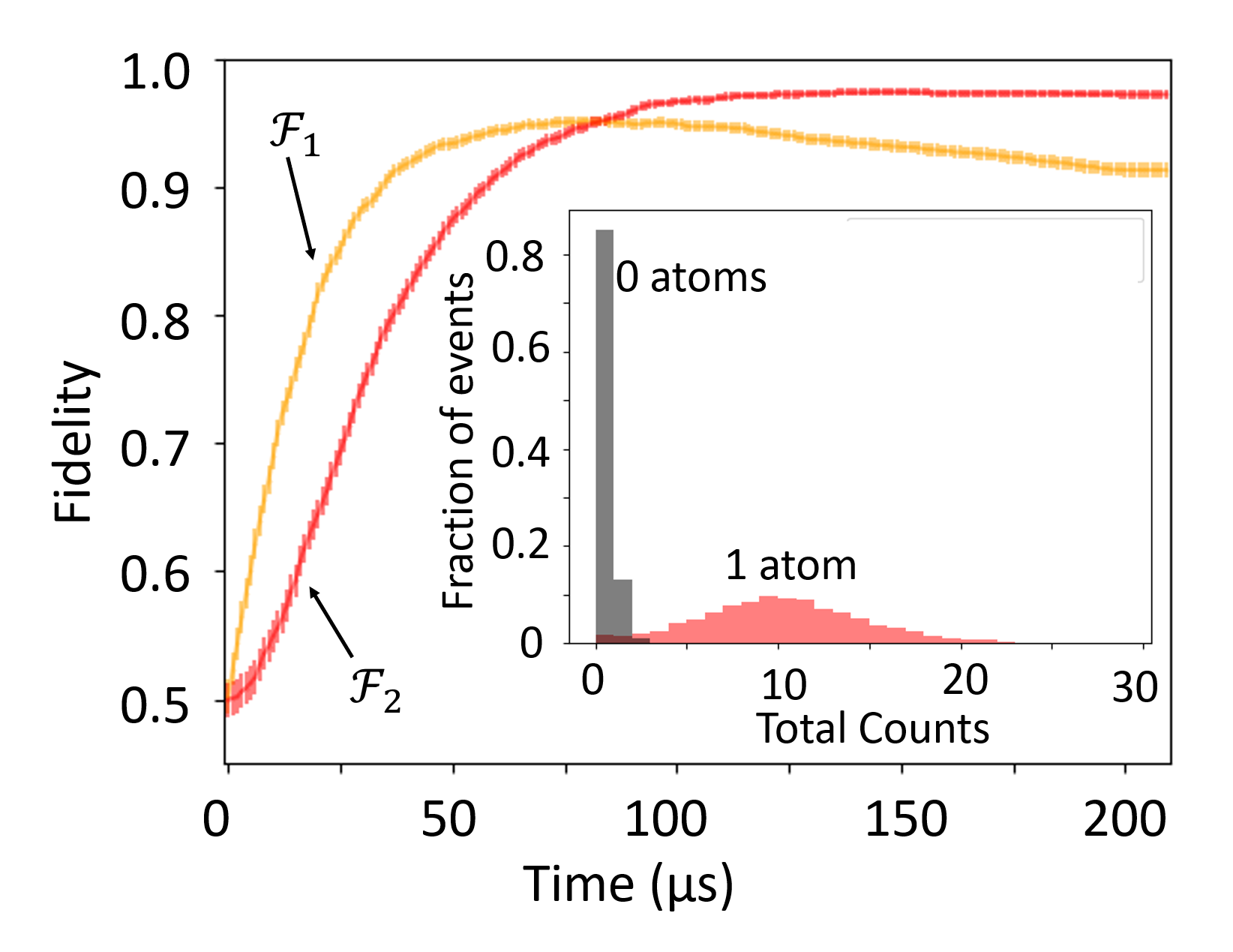} \end{center}
	\caption{$\mathcal{F}_{1}$ (yellow) and $\mathcal{F}_{2}$ (red) are determined for each $\mu$s during the 200 $\mu$s-long probe pulse. The inset shows the histograms of events used to generate the fidelity curves. The inset data is taken at 200 $\mu$s. }\label{fig:fidtimes}
\end{figure}

We find that $\mathcal{F}_{n_{thresh}}$ is optimized when the readout-beam frequency is  $\nu_{23}+40$ MHz, which is different from the frequency that the maximizes the total fluorescence ($\nu_{23}+46$ MHz). As seen in Fig.~\ref{fig:fidtimes}, the fidelity obtained using $n_{thresh}=1$, denoted by $\mathcal{F}_{1}$, has a maximum value of $95.0\pm0.3\%$ at a measurement time of $84\pm6$ $\mu$s.  For thresholding on two photons ($n_{thresh}=2$), denoted by $\mathcal{F}_{2}$, a maximum fidelity of $97.6\pm0.2\%$ is obtained for a measurement time of $160\pm20$ $\mu$s. This is the fastest non-destructive quantum state readout reported for a neutral atom trapped in a free-space ODT. We find that using more than two photons to classify the atom as bright or dark does not improve the fidelity, in agreement with Ref. \cite{BrowaeysND}. This readout fidelity was achieved with a probe beam power of 200 $\mu$W.

The curves in Fig. \ref{fig:fidtimes} represent several experimental runs totaling 3,583 experiments in which the atom is prepared in the bright state and 3,550 experiments in which the atom is prepared in the dark state. The data is post-selected for those events where an atom is retained in the trap. On average, the atom is retained in the trap after readout in $97.1\pm0.1\%$ of the experiments. This retention number is not corrected for background-induced losses.  

Examining the photon arrival times shows that the atom's scattering rate decreases during the measurement. One known cause of this loss is off-resonant pumping (ORP) causing the atom to drop into the $F=1$ ground state \cite{NoekOL, SaffmanND}. To quantify the scattering rate loss, we turn to a model developed by the ion trap community to describe fluorescence detection \cite{NoekOL, Crain}. The model assumes that the atom starts in an initial state with scattering rate $R_{i}$. During the probe pulse, there is the possibility that the atom will off-resonantly pump to another state with rate $R_{f}$ and the loss of the atom from the bright state happens with probability $R_{l}$. The total probability of an atom scattering $n$ photons in time $t$ is given by 
\begin{equation}
\label{eq:Ptot}
\begin{aligned}
& P_{Tot}(n;t,R_{i},R_{f},R_{l})= \\ & e^{-(R_{l}t)}P_{ph}(n;R_{i}t)+P_{l}(n;t,R_{i},R_{f},R_{l}),
\end{aligned}
\end{equation}
where the first term is the probability that the atom does not transition during time $t$ and all of the photons are scattered when the atom is in state $i$. This non-transition probability is given by
\begin{equation}
\label{eq:PoissC}
P_{ph}(n;t,R_{i})=e^{-R_{i}t}\frac{(R_{i}t)^{n}}{n!}.
\end{equation}
The second term is the probability of scattering $n$ photons while undergoing a transition and is given by
\begin{equation}
\label{eq:PtrFullC}
\begin{aligned}
& P_{l}(n;t,R_{i},R_{f},R_{l})=\sum_{k=0}^{n}R_{l}e^{-R_{f}t}\int_{0}^{t}d\tau \frac{(R_{i}\tau)^{k}}{k!} \\  & \frac{(R_{f}t-R_{f}\tau)^{(n-k)}}{(n-k)!}e^{-(R_{i}-R_{f}+R_{l})\tau}.
\end{aligned}
\end{equation}
where the transition occurs at time $\tau$ within the interval 0 to $t$ and $k$ photons are scattered from state $i$ and the remainder from state $f$. The full derivation of the model is given in \cite{Shea}.

For an atom prepared in the bright state $b$ the atom scatters photons at a rate of $R_{0}$ and the detector collects photons at the rate $\eta R_{0} + R_{bg}$, where $\eta$ is the detection efficiency of the system and $R_{bg}$ is the rate of background counts entering the detector. During the readout, there is the possibility that the atom will be lost to the dark state through off-resonant pumping at rate $R_{l}$. We assume that the atom does not transition back to the bright state $b$ after it transitions to the dark state $d$. This assumption is reasonable for the short readout times considered here. Using Eq. \ref{eq:Ptot}, we find that the probability that the bright-state atom will scatter $n$ photons in time $t$ is given by
\begin{equation}
\label{eq:Pbright}
\begin{aligned}
& P_{Tot;b}(n;t)=e^{-(\eta R_{0}+R_{bg}+R_{l})t}\frac{(\eta R_{0}+R_{bg})^{n}t^{n}}{n!}+ \\ &  \left(\frac{R_{l}e^{-R_{bg}t}}{\eta R_{0}+R_{l} }\right)\left(\frac{\eta R_{0}}{\eta R_{0}+R_{l} }\right)^{n} \\ &  \times\left[\sum_{k=0}^{n}\frac{(\eta R_{0}+R_{l} )^{k}(R_{bg}t)^{k}}{k!(\eta R_{0})^{k}} - e^{-(\eta R_{0}+R_{l})t} \right. \\ & \left. \times \sum_{k=0}^{n}\frac{(\eta R_{0}+R_{l} )^{k}(\eta R_{0}+R_{bg})^{k}t^{k}}{k!(\eta R_{0})^{k}}\right] .
\end{aligned}
\end{equation}
The bright state error is defined as the probability that the atom has scattered fewer than $n_{thresh}$ photons during the collection time and is given by
\begin{equation}
\label{eq:Berror}
E_{b}(t)=\sum_{k=0}^{n_{thresh}-1}P_{Tot;b}(k;t).
\end{equation}

\begin{figure}
	\begin{center} \includegraphics[width=3.25 in]{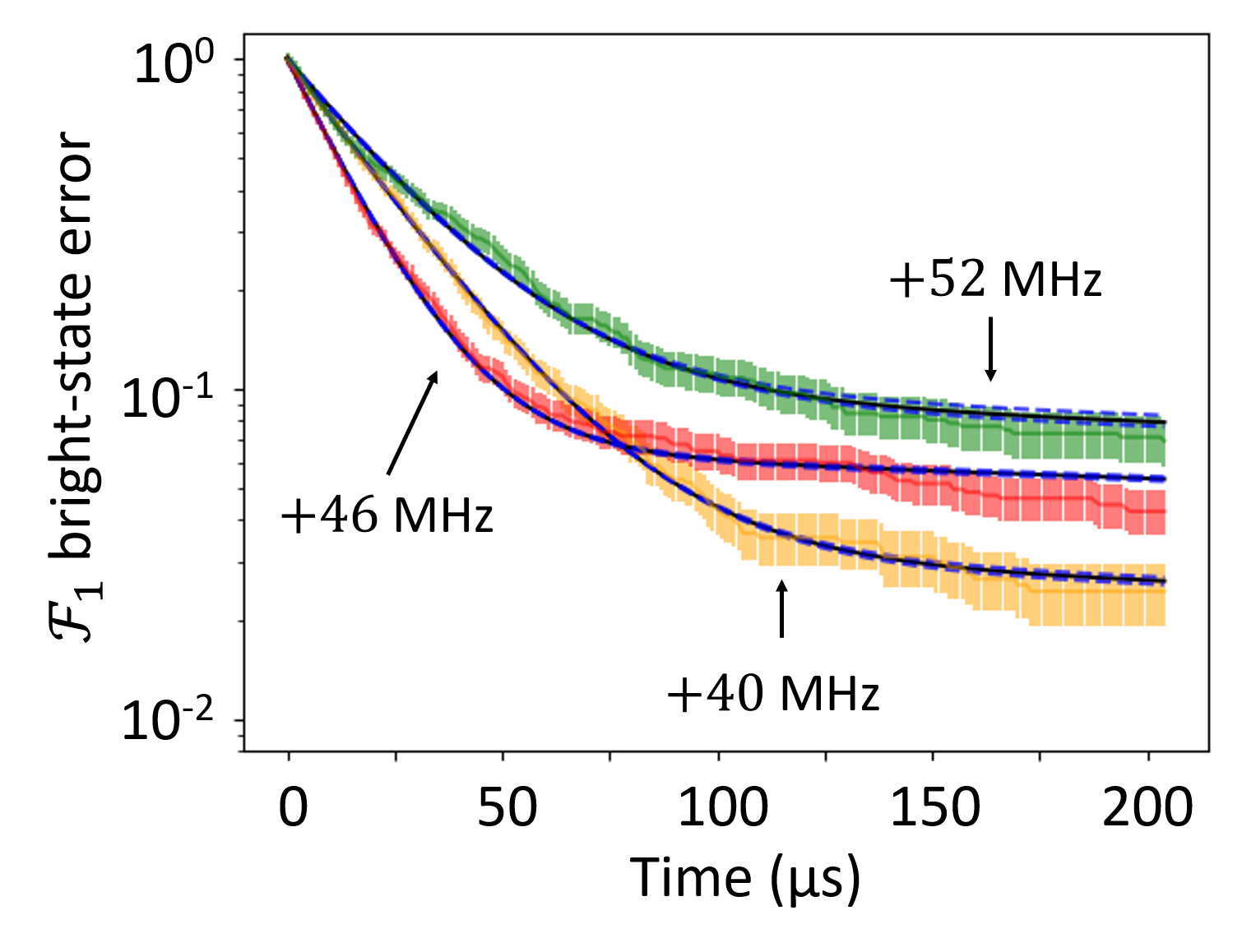} \end{center}
	\caption{$\mathcal{F}_{1}$ bright-state error rates for three probe frequencies. The solids lines are the fits to the protocol model with the dashed lines representing the 95\% confidence levels. }\label{fig:3errors}
\end{figure}

Naively, one would expect off-resonant pumping to be largest for frequencies tuned below the shifted resonance because such frequencies are closer to the $F'=2$ excited state through which ORP occurs \cite{MeschedeDFF}. To test this, we prepare the atom in the bright state and perform state readout at three different probe frequencies. One readout is performed with the probe frequency set to the frequency at which the atomic fluorescence peaks, $+46$ MHz shifted from the untrapped atomic resonance. Another data set is taken at $+40$ MHz of the untrapped atomic resonance, where we find the peak readout fidelity, and a third at $+52$ MHz of the untrapped resonance. For each probe frequency, we plot the bright-state error based on $n_{thresh}=1$ and fit the data to the model of Eq. \ref{eq:Pbright}. The results are shown in Fig. \ref{fig:3errors}. We measure $R_{bg}$ for each data set using the photons counted for those trials in which an atom is not trapped. We also measure the detection efficiency $\eta$ of our system using the saturation method detailed in \cite{Shea} that is an extension of that reported in Ref. \cite{Crain}. $R_{0}$ and $R_{l}$ are left as free parameters in the model. 

The results of the fitting procedure are given in Tbl. \ref{tbl:errorfits}. The background rate $R_{bg}$ is relatively consistent across all three data sets, so the shape of the curve is dependent on the interplay between $R_{0}$ and $R_{l}$. Initially, $\epsilon_{B}$ falls quickly at a rate governed by $R_{0}$ before leveling off due to the loss in counts caused by $R_{l}$. Thus, the lowest possible bright-state error is achieved for the largest ratio of $R_{0}$ and $R_{l}$. This occurs for the $+40$ MHz readout beam, mainly because $R_{l}$ is smallest for this case. This is opposite to our intuition about ORP, because this readout-beam frequency is tuned closest to the $F'=2$ excited state. 
\begin{table}
	\caption{\label{tbl:errorfits} The values for $\eta R_{0}$ and $R_{l}$ at each probe frequency extracted from the protocol model fit to experimental data. The measured $R_{bg}$ are included for completeness. The values are given in kilo-counts per second (kcps)}
	\begin{center}
		\begin{tabular}{|c|c|c|c|c|} \hline
			{\bf Probe f} & {\bf $R_{bg}$ (kcps)} & {\bf $\eta R_{0}$ (kcps)} &  {\bf $R_{l}$ (kcps)}&  {\bf $\eta R_{0}/R_{l}$} \\ \hline
			+40 MHz & 1.05 & $39.4\pm0.2$  &  $1.31\pm0.04$ &30.08\\ \hline
			+46 MHz & 1.13 & $58.7\pm0.5$ &  $4.1\pm0.1$ &14.3 \\ \hline
			+52 MHz & 1.12 &  $33.6\pm0.3$ & $3.63\pm0.1$ & 8.2\\ \hline
		\end{tabular}
	\end{center}
\end{table} 

We believe this discrepancy between intuition and measurement is due to multiple factors contributing to the loss of count rate captured by $R_{l}$. In addition to ORP, the atom experiences heating during quantum state readout that can cause a change in scattering rate. As the atom heats, it samples a larger range of trap depths and is farther detuned from the probe beam, lowering the scattering rate \cite{Shea}. Heating is known to be important for single-atom traps \cite{MeschedeDFF,KurtseiferHeat} and may contribute to the increase in $R_{l}$ on and above resonance, where heating is known to be worse \cite{Metcalf}.

\begin{figure}
	\begin{center} \includegraphics[width=3.25 in]{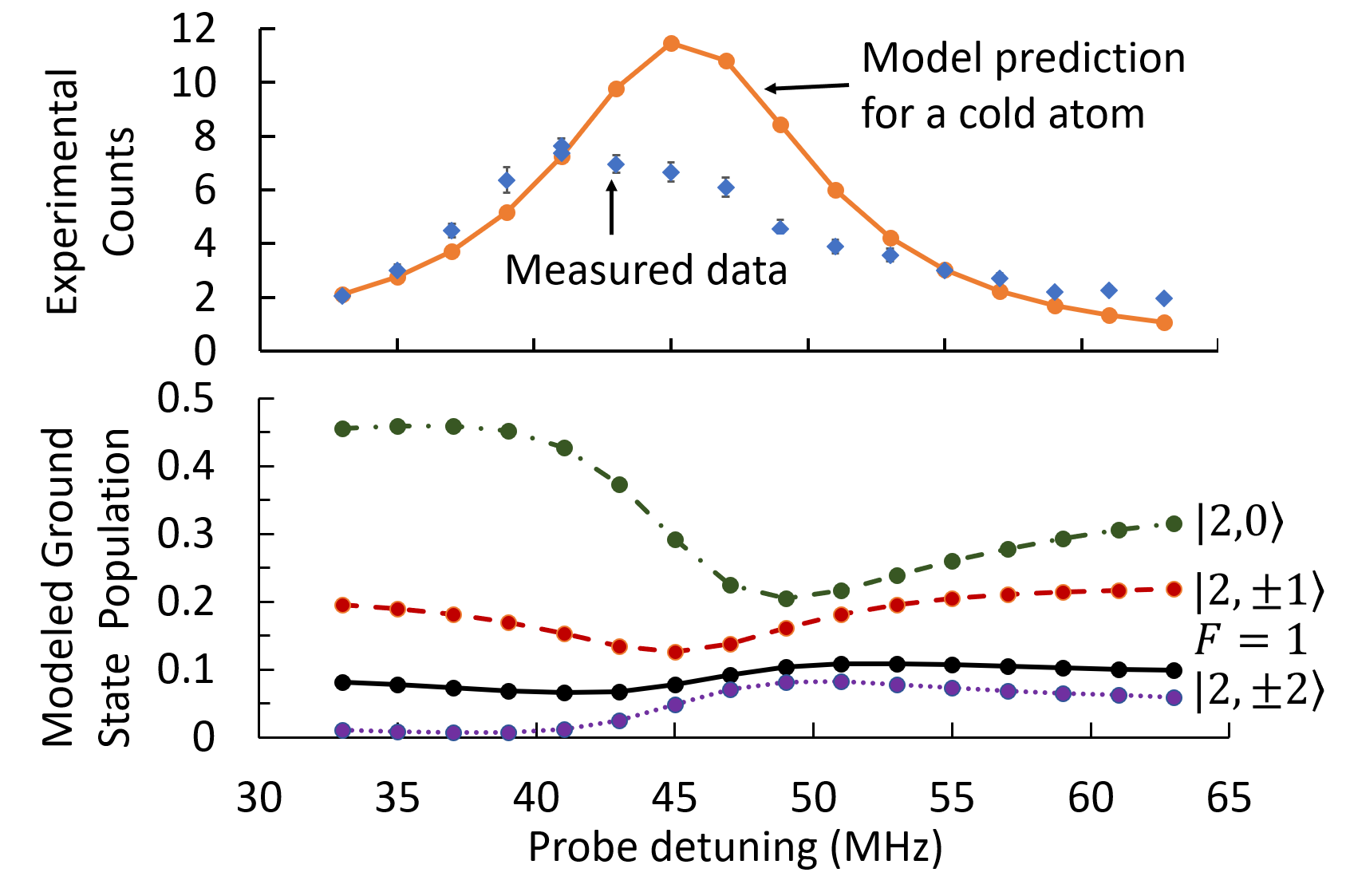} \end{center}
	\caption{{\em Top Panel} Compared to the prediction for a cold atom (orange) the atomic fluorescence (blue) is suppressed at higher detunings. {\em Bottom Panel}  Rate-equation model predictions for the ground state atomic sublevel populations for various probe beams.}\label{fig:heating}
\end{figure}

We look for evidence of heating in our system by comparing the frequency-dependence of the atomic fluorescence during quantum state readout to that seen during atom detection, when cooling is present. The data is plotted in the top panel of Fig. \ref{fig:heating}. The solid line is a rate equation model of the system that does not include heating. We see that it accurately predicts atomic behavior when the atom remains cold, but fails during quantum state readout. Near and above resonance, we see that the atomic fluorescence is suppressed during the quantum state readout. This is the behavior captured by $R_{l}$. The peak fluorescence during readout occurs at a detuning near the half-width at half-maximum of the cooled atom's broadened resonance, the location known to minimize Doppler heating \cite{SaffmanND}. This supports the hypothesis that heating contributes to $R_{l}$. Time-resolved fluorescence detection of single-atoms, such as that demonstrated here, could be a useful tool for further investigating such heating effects. Incorporating such effects into a rate-equation model of the atom-probe system would be a natural extension of this work.

Heating is likely a significant cause of $R_{l}$, but off-resonant pumping is known to contribute as well, and we would still expect it to be worse below peak-resonance. Our rate equation model contains the mechanisms for off-resonant pumping and can shed light on this physical process. ORP can be caused by transitions through the $F'=2$ or the $F'=1$ excited states. Both types of transitions are captured by our model but we find that transitions through $F'=2$ are dominant due to the additional detuning of the probe beam from the $F'=1$ state.

The ground state populations predicted by the rate equation model are shown in the bottom panel of Fig. \ref{fig:heating}. We see that the $F=1$ population, which is proportional to the amount of ORP, peaks above the location of the shifted resonance. This result can partially explain the higher values of $R_{l}$ above resonance and can be understood by considering the $F=2$ sublevel populations shown in Fig. \ref{fig:heating}. The atomic population always preferentially accrues in the $m_{F}=0$ state due to the Clebsch-Gordan coefficients of the relevant transitions. This state is relatively protected from ORP because the $|2,0\rangle \rightarrow |2,0\rangle$ transition is quantum mechanically forbidden. At frequencies below the $|2,0\rangle \rightarrow |3,0\rangle$ transition, the protected $|2,0\rangle$ state dominates the population, suppressing transfer of population to $F=1$. At higher frequencies, however, relatively more atomic population accrues in the $|2,\pm1\rangle$ and $|2,\pm2\rangle$ states. These states are not protected from ORP in the same way, so the amount of population in $F=1$ increases. Thus, we see that the use of $\pi$-polarized light tuned at or below the atomic resonance exploits the quantum mechanical selection rules to suppress ORP.  

We have shown that state readout fidelity depends on three parameters: $R_{bg}$, $\eta R_{0}$, and $R_{l}$. This is a powerful model for optimizing non-destructive state detection. Decreasing $R_{bg}$ lowers the dark-state error and improves fidelity. The background rate is around 700 cps for the data presented in Fig. \ref{fig:fidtimes}. The largest source of background scatter in our experiment is stray scatter from the probe beam. This can be decreased by focusing the probe beam, therefore decreasing the power needed to reach the desired intensity \cite{BrowaeysND} and by using a fiber with a smaller collection core to improve spatial filtering. Both of these techniques greatly increase the alignment difficulty of the system and were beyond the scope of this work.

The bright-state error can be reduced by increasing $\eta R_{0}$ and decreasing $R_{l}$. In our system, we measure a total detection efficiency of $0.96\%$. Aberrations in the imaged fluorescence are the main source of loss in our collection path. Improved alignment, which is not experimentally trivial, should increase the collection efficiency. Using the readout model described here, we estimate that increasing $\eta$ by a factor of two will yield a peak fidelity of $F_{2}=98.8\%$ in a $\sim$75-$\mu$s-long detection time. Clearly, linearly-polarized readout light offers a promising route to fast, nondestructive quantum-state readout.

The final parameter of interest, $R_{l}$ is less straightforward to modify, as it likely depends on both ORP and heating of the atom. A demonstrated method to suppress ORP is to prepare the atom into one of the $m_{F}=\pm2$ states and use circularly polarized readout light on the $|2,2\rangle \rightarrow |3,3\rangle$ cycling transition. This has been shown to achieve readout fidelities of $>98\%$  \cite{BrowaeysND,SaffmanND,MeschedeND} but requires a more complicated state preparation scheme than that used here. Furthermore, pumping the atom into the stretched state is relatively slow because of the differential shifts caused by the ODT. Linearly-polarized readout schemes like that presented here and in \cite{ChapmanND} are consistently an order of magnitude faster than those utilizing circularly-polarized light \cite{BrowaeysND,SaffmanND,MeschedeND}.

In conclusion, we demonstrate the fastest non-destructive quantum-state readout yet reported for neutral atoms trapped in free-space. Using linearly-polarized light and time-tagging the detected photons, we investigate the time-dependence of the atomic scattering rate during the probe pulse and identify a mechanism for protecting the atom from ORP by tuning the readout light frequency to just below resonance. We adapt a model of the readout process from the ion trap community and couple that with a rate equation model of the atomic populations to gain insight into the atom's behavior during readout. These techniques can be used to further understand the interaction of single, neutral atoms with near-resonant laser light and off-resonant trapping light.

\vspace{5mm}

We gratefully acknowledge the support of the US Army Research Office cooperative agreement Award No. W911NF-15-2-0047.


\bibliography{MSheaReadoutBibl_PRA}

\end{document}